\documentclass{an}
\usepackage{graphicx}
\usepackage{times}
\usepackage{fancyhdr}
\begin{document}

\title{Old metal-rich globular cluster populations: Peak color
and peak metallicity trends with mass of host spheroids}

\author{V.V. Kravtsov\inst{1,2}}
\institute{ Instituto de Astronom\'ia, Universidad Cat\'olica del
Norte, Avenida Angamos 0610, Antofagasta, Chile
\and
Sternberg Astronomical Institute, University Avenue 13, 119899
Moscow, Russia}

\date{Received $<$date$>$;
accepted $<$date$>$;
published online $<$date$>$}

\abstract{ We address the problem of the factors contributing to a
peak color trend of old metal-rich globular cluster (MRGC)
populations with mass of their hosts, early-type galaxies and
spheroidal subsystems of spiral ones (spheroids). The color-mass
trend is often converted to a metallicity-mass trend under the
assumption that age effects are small or negligible. While direct
estimates of the ages of MRGC populations neither can rule out nor
reliably support the populations' age trend, key data on timing of
the formation of spheroids and other indirect evidence imply it in
the sense: the more massive spheroid the older on average its MRGC
population. We show that the contribution of an allowable age trend
of the MRGC populations to their peak color trend can achieve up to
$\sim$50\% or so. In this event the comparable value of the color
trend, $\sim$30\%, is due to alpha-element ratio systematic
variations of the order of $\Delta$[$\alpha$/Fe] $\approx$ 0.1 to
0.2 dex because of a correlation between the [$\alpha$/Fe] ratios
and age. Hence a systematic variation of exactly [Fe/H] ratios may
turn out to be less significant among the contributors, and its
range many times lower, i.e. of the order of $\Delta$[Fe/H]$\sim$0.1
or even none, than the corresponding range deduced by assuming no
age trend. 
}

\correspondence{vkravtsov@ucn.cl}

\maketitle

\section{Introduction}
Knowledge of the basic characteristics of old globular cluster (GC)
populations in spheroidal subsystems of spiral galaxies and in
early-type galaxies (thereafter spheroids) of wide range of their
mass is important for adequate understanding of early galactic
history as well as of processes (factors) responsible for the
formation of the populations. Among these characteristics, such as
kinematic, spatial distribution and age, are peak metallicities of
the GC populations. They provide us with valuable information on the
relationships between the formation of the populations, on the one
hand, and star formation (SF) in galaxies and chemical enrichment
processes in them, on the other hand. In spite of essential progress
achieved for the past decade in obtaining new significant data and
in studying these relationships, some details remain unclear or do
not receive sufficient attention. Among them seem to be those
related to the dependence of metallicity of the old GC populations
on mass of parent spheroids.

The majority of spheroids are known to have formed two populations
of old GCs, metal-poor (MP) and  metal-rich (MR) ones. While the
MPGCs are virtually ubiquitous in spheroids of nearly any mass
(except for least massive dwarf spheroidal galaxies), the MRGCs are,
as a rule, systematically very scanty or absent at all among the
least massive hosts, and the MRGCs-to-MPGCs ratio increases
systematically with increasing mass of the host spheroids (Peng et
al. 2006). The bimodal nature of the metallicity distribution of the
old GCs in galaxies is primarily reflected in the respective
bimodality of the clusters' color distribution exhibiting two more
or less distinct peaks. Transformations from colors of GCs to their
metallicities are based on a number of currently available
color-metallicity relations. These relations have been
obtained by different groups relying on most reliable data on
metallicities and colors, namely the ($V-I$) and ($g-z$) ones of the
Galactic GCs and their extragalactic counterparts (e.g.,
Kissler-Patig et al. 1998; Kundu \& Whitmore 1998; Peng et al.
2006). Note that the MPGCs are beyond the scope of the present paper
which is devoted to old MRGC populations and to problems related to
their peak metallicities and mean ages.

Peak colors of MRGC populations are in turn observed to correlate
with mass (luminosity, velocity dispersion) of parent spheroids, in
the sense: the more massive spheroid, the redder peak color of its
MRGC population (Forbes \& Forte 2001; Larsen et al. 2001; Kundu \&
Whitmore 2001; Peng et al. 2006; among others). Although the MPGCs
are not discussed here, it is useful to note that they also exhibit
a peak color-luminosity trend similar to but less significant than
that observed for the MRGCs (e.g., Larsen et al. 2001; Strader et
al. 2006; Peng et al. 2006). At first glance, these color variations
seem to be of the same nature as the bimodality of old GC color
distribution, i.e. mainly or fully due to the respective metallicity
variations. Currently available color-metallicity relations used for
the corresponding transformations  assume MRGC populations in
spheroids of different mass to be more or less coeval, with no
(essential) trend of their age with parent galaxy mass. This
approximation is due to (1) the lack of firm direct evidence for the
dependence of the ages of MRGC populations on galaxy mass, and (2)
the lack of suitable empirical calibrators of the color-metallicity
relation for other than old ages. However, it is clear that it is
only first approximation to the problem under consideration.
Resolving it can have a number of important consequences,
particularly for adequate understanding of the formation of MRGCs.
Here, we address the problem of the age trend among MRGC populations
by relying principally on an available body of indirect evidence.
Hence we attempt to estimate plausible constraints that may be
imposed on the contribution of metallicity to a color trend. Also,
we briefly discuss their implications for the MRGC formation.

In Sect.~\ref{spheroids} we briefly summarize the basic results on
timing of the formation of spheroids, obtained from studies of both
high redshift galaxies and nearby ones. The conservative upper limit
of possible systematic age difference between MRGC populations in
spheroids of different mass is considered in Sect.~\ref{mrgcages}.
Its contribution to a peak color trend of MRGC populations with mass
of the host spheroids is estimated in Sect.~\ref{contrib}.
Sect.~\ref{sumrem} contains summary and concluding remarks.

\section{Timing of the formation of spheroids}
\label{spheroids}

The key results obtained to date on early stages and timing of the
formation and evolution of spheroids provide a valuable
observational information for more adequate understanding of the
same aspects concerning old MRGC populations.

Approximately linear relation between masses of supermassive black
holes (SBHs) and host stellar bulges (Magorrian et al. 1998;
Gebhardt et al. 2000; Graham et al. 2001) implies that the formation
of these components of spheroids must be related and that the growth
of SBHs and massive bulges occurred simultaneously. Therefore, the
bulk of stars in spheroids formed at the same time as their SBHs
acquired most of their mass (e.g., Page et al. 2001; Granato et al.
2001; Shields et al. 2003). SF activity and its respective impact on
accretion onto the SMBs in early spheroids are tightly bound to the
basic characteristics of high redshift QSOs, as well as to the
properties of the QSO luminosity function and its redshift
evolution. Haiman, Ciotti \& Ostriker (2004) argue that the redshift
evolution of the QSO emissivity and of the SF history in spheroids
should be approximately parallel. According to Granato et al.
(2001), the evolution of QSOs and galaxies can be well understood if
we accept that spheroids of different mass form the bulk of their
stars on different timescales: the more massive spheroid (and its
SBH), the shorter timescale. Recent observations are in good
agreement with this conclusion. Correlations have been found between
velocity dispersion and age (e.g., Caldwell, Rose \& Concannon
2003), between [$\alpha$/Fe] and velocity dispersion, as well as
between the alpha-element abundance ratios and mean ages (Thomas,
Maraston \& Bender 2002) of early-type galaxies. Hence one deduces
that more massive galaxies had shorter timescales of their star
formation. Indeed, data on high redshift objects, around z$\sim$6,
reveal the very massive SBHs and galaxies with high SF rate to form
in the early Universe (Bertoldi \& Cox 2002; Bunker et al. 2003;
Willott, McLure \& Jarvis 2003; Freudling, Corbin \& Korista 2003).
At the same time, the timescales may differ, to a certain extent,
for galaxies of the same mass forming in high and low density
environments (Thomas et al. 2005).

The deduced timescale of the formation of the bulk of stars in the
most massive spheroids, with velocity dispersion around or exceeding
$\sigma\approx$ 300 km s$^{-1}$, is short as compared to a Hubble
time. In the above-cited publications as well as in a number of
other ones devoted to the formation of high redshift galaxies, the
timescale is estimated to be of order or even less than 1 Gyr since
the Big Bang. In turn, spheroids of systematically lower mass
achieve the epochs of their highest SF rate systematically later.
According to the estimates by Thomas et al. (2002), for example, an
age difference between these epochs in the very massive spheroids
and in those with $\sigma\approx$ 80 km s$^{-1}$ is around 10 Gyr.

It has to be noted, however, that the ages under consideration, i.e.
ones estimated for stellar populations confined in spheroids, are
often luminosity weighted. They can significantly be affected by
stars formed in last essential burst of star formation in a galaxy
rather than by the bulk of the galaxy's populations formed (much)
earlier.

\section{Implied age difference between MRGC populations in spheroids of different mass}
\label{mrgcages}

Observations of galaxies in the nearby Universe unambiguously reveal
a tight relationship between high SF rate and formation of as
massive star clusters as GCs (e.g., Larsen \& Richtler 2000; Larsen
2000, 2002). Similarly, evidence for the formation of both MRGCs and
metal-rich stars in early spheroids in the same star formation
events, with similar ages and metallicities has been presented by
Harris, Harris \& Poole (1999), Durrell, Harris \& Pritchet (2001),
Forbes \& Forte (2001) and Forbes (2002).

The estimated epochs of the highest SF activity in spheroids of wide
mass range imply that the bulk of MRGCs populating the most massive
of the hosts are very old and nearly coeval with MPGCs because their
mean age difference is expected to fall between a few hundred Myr
and 1 Gyr. Moreover, MRGC populations in lower mass hosts are likely
systematically younger than in the higher mass ones, similarly to
the hosts' field stars. In other words, the conclusions achieved to
date on the timescale of the formation of spheroids with different
mass provide important implication for the MRGCs populating the
hosts: the more massive spheroid (or the higher its velocity
dispersion), the older on average its MRGC population. At the same
time, it cannot be excluded that GCs may be somewhat older than the
field stellar populations in the same elliptical galaxies (Puzia,
Kissler-Patig \& Gougfrooij 2006). In other words, age trends of
MRGCs and of the corresponding metal-rich field stellar populations
may be somewhat different.

\begin{figure}
\resizebox{\hsize}{!}
{\includegraphics[angle=-90]{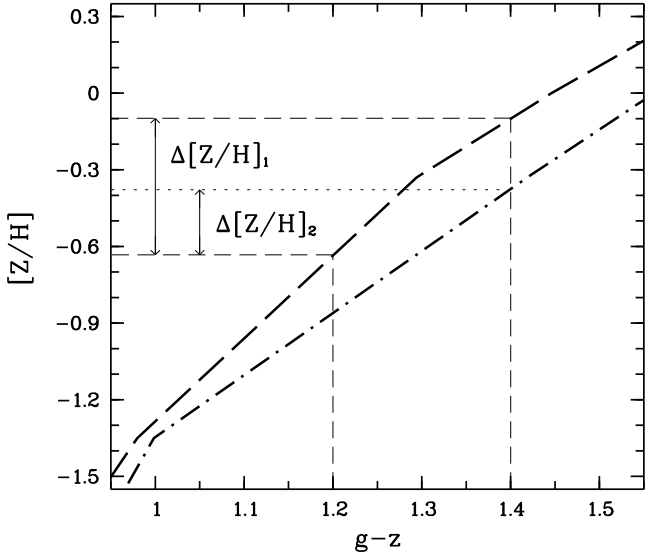}}
\caption{The effect
of possible systematic age difference among MRGC populations
belonging to spheroids of different mass on a trend of their [Z/H]
peak metallicity as a function of their ($g-z$) peak color;
color-metallicity relations for ages of 9 Gyr and 14 Gyr are
shown by long-dashed line and dash-dotted line, respectively, using
evolutionary population synthesis models by Maraston (2005); for more
details and explanations see the text.} \label{fig:1}
\end{figure}

Due to their limited accuracy of a few Gyr at best, the direct
estimates of the ages of MRGCs in spheroids of different mass
neither can rule out such an age trend nor reliably support it at
least within 4-6 Gyr since the Big Bang. As a rule, such estimates
are made for the MRGCs belonging to spheroids of higher mass, with
stellar mass M$_s >$ 10$^{10}$M$_\odot$, in which the clusters can
be isolated in sufficient number (tens). The MRGC populations are
typically concluded to be either older than 8-10 Gyr or as old as
MPGCs (see, for example, Puzia et al. 1999; Larsen et al. 2002;
Puzia et al. 2002; Beasley et al. 2004; Strader et al. 2005; Pierce
et al. 2006). It is now difficult to reliably conclude on such a
trend and on its value. However, its conservative upper limit of
$\sim$5 Gyr, corresponding to the mentioned uncertainty of MRGC ages
in spheroids with stellar mass M$_s >$ 10$^{10}$M$_\odot$, seems to
be quite realistic. Indeed, Puzia et al. (2005) estimate that up to
one third of GCs in early-type galaxies have ages in the range 5-10
Gyr. It has to be noted, however, that this result is based on very
limited sample of GCs, and it is not clear whether there is any
trend of mean age of the MRGCs with galaxy mass. Among other data
that might count in favor of such a trend we refer to those obtained
recently on GCs in the giant elliptical galaxy M87, on the one hand,
and in dwarf ellipticals, satellites of the Andromeda galaxy, on the
other hand. Sharina, Afanasiev \& Puzia (2006) find three MRGCs (
[Z/H]$\geq-$0.8 dex) in NGC 185 and NGC 205. All of them are of
apparently younger ages ($\leq$ 7 Gyr) than their Galactic
counterparts. In turn Sohn et. (2006) have found that GCs in M87 are
notably bluer in the (FUV$-V$) color than those in the Galaxy,
including MRGCs. Older ages of GCs in M87 is one of the basic
factors that can be responsible for this color difference.

It is worth of noting that the difference in the slopes of the
color-metallicity relations (namely in their metal-rich range)
obtained by Kundu \& Whitmore (1998) and Kissler-Patig et al. (1998)
relying on the data on GCs of the Milky Way and the cD galaxy NGC
1399, respectively, is in agreement with older ages of the MRGCs
belonging to NGC 1399. It is interesting that the peak metallicity
deduced from peak color of the MRGCs in NGC 1399, using just the
latter color-metallicity relation based on spectroscopy of a
sub-sample of the MRGCs from the same galaxy, is [Fe/H]$=-$0.6. The
same relation gives [Fe/H]$=-$0.5 for MRGCs of another very massive
galaxy, the giant elliptical M87 (Kissler-Patig et al. 1998). By
calibrating the ($B-V$) colors of GCs in the spiral galaxy M81 with
spectroscopically based metallicities of a sub-sample of the
galaxy's clusters, Ma et al. (2005) obtain [Fe/H]$=-$0.53 for the
M81 MRGC population. Therefore, these peak values of [Fe/H] are
indistinguishable within the errors from each other and from peak
metallicity of the Galactic MRGCs. In other words, in the case when
peak metallicities of MRGC populations, belonging to galaxies of
different mass, are derived either from spectroscopy or from
spectroscopically based color-metallicity relations obtained for the
same parent galaxies, the deduced peak values of [Fe/H] converge.
Moreover, the difference between these peak values becomes smaller
as compared to that derived only from peak colors by means of any
single color-metallicity relation.

\section{Effect of an age trend on the deduced variation of peak metallicity}
\label{contrib}

To estimate a contribution of the assumed age trend to color
trend of MRGC populations and consequently the deduced
variation of peak metallicity, we combine (1) color-metallicity
relations for appropriate ages, based on evolutionary
population synthesis models, with (2) observational data on real color
range corresponding to systematic spread of peak colors of MRGC
populations belonging to spheroids of the mass range under
consideration. For this purpose we used the latest data of Peng et al.
(2006) on the dependence of the ($g-z$) peak colors of the MRGC
populations on luminosity (stellar mass) of their parent early-type
galaxies belonging to the Virgo cluster. For galaxies with stellar
mass M$_s >$ 10$^{10}$M$_\odot$, peak colors of their MRGC
populations are found to fall between 1.2$\leq(g-z)\leq$1.4.

Fig.~\ref{fig:1} demonstrates the effect of possible systematic age
difference among MRGC populations on a deduced trend of their [Z/H]
peak metallicity as a function of their ($g-z$) peak color. In order
to estimate the effect, one of the currently available sets of
evolutionary population synthesis models is used. Specifically, we
rely on recently developed models by Maraston (2005). Presented are
those of them whose age differs by 5 Gyr, namely models with age of
9 Gyr and 14 Gyr, and in the metallicity range under consideration.
In Fig.~\ref{fig:1} they are connected by long-dashed line and
dash-dotted line, respectively. These are models with Salpeter
initial mass function and with blue horizontal branch at the two
lowest metallicities, [Z/H]=$-2.25$ and [Z/H]=$-1.35$. The color
range of $\Delta(g-z)$=0.2 mag is translated to metallicity range of
$\Delta$[Z/H]$_1\approx$ 0.5 dex if color-metallicity relation of
the same age is used: in our example it corresponds to the models
with age of 9 Gyr (long-dashed line). Note that approximately the
same metallicity range, but at somewhat lower mean metallicity, is
deduced using models with age of 14 Gyr (dash-dotted line). However,
twice as lower metallicity range of peak metallicities,
$\Delta$[Z/H]$_2 \approx$ 0.25 dex, would be deduced if the MRGC
populations having the color $(g-z)$=1.2 and belonging to the
lower-mass galaxies (M$_s \approx$ 10$^{10}$M$_\odot$) were 9 Gyr
old whereas their counterparts populating galaxies of highest mass
and exhibiting redder peak color, $(g-z)$=1.4, were 5 Gyr older. It
is evident that at a given metallicity, say [Z/H]$\sim -0.3$ to
$-0.4$ dex, according to the models used, the effect of the
discussed age difference among MRGC populations in galaxies of
different mass may achieve around 0.12 mag in the color ($g-z$).

Virtually the same is valid regarding a trend of the ($V-I$) color.
We mention here the results of Larsen et al. (2001) obtained from a
quite large homogeneous set of data on GCs in early-type galaxies
spanning ranges of absolute $B$-magnitude, $M_B$, and
central$\sigma$: $\Delta M_B\approx 3.5$ mag and $\Delta$log$\sigma
\approx 0.6$ dex, respectively. These variations of the galaxy
characteristics approximately correspond to the above-considered
range of galaxy mass. The least-squares fits to the data plotted as
relations between ($V-I$) peak colors of MRGC populations and $M_B$
or log$\sigma$ show the difference of $\Delta(V-I)\sim$ 0.08-0.10
mag corresponding to the given ranges of the galaxy characteristics.
Similarly, approximately $4-5$ Gyr of the systematic age difference
among MRGC populations belonging to spheroids of the mentioned mass
range can account for around 50\% of the color trend.

It has to be noted that in the case of an age trend of MRGC
populations (in galaxies with M$_s >$ 10$^{10}$M$_\odot$) of the
order of 5 Gyr or more, it is a trend of their [$\alpha$/Fe] ratios
with age that may be the mainly or even the only responsible for the
peak metallicity trend that corresponds to the rest of peak color
trend (after taking into account the effect of age). Indeed, by
analyzing [$\alpha$/Fe], metallicity, and age distributions of GCs
in elliptical, lenticular, and spiral galaxies, Puzia et al. (2006)
find (tentative) evidence for an age$-$[$\alpha$/Fe] correlation for
GCs at least in elliptical galaxies. This correlation is expressed
in increasing values of [$\alpha$/Fe] ratios with age of GCs. By
relying on these data we estimate that the ratios may change by at
least $\Delta$[$\alpha$/Fe]$\sim 0.1 - 0.2$ dex for the accepted
period of 5 Gyr. For definiteness we accept 0.15 dex. From relation
between total metallicity, [Z/H], and its constituents, [Fe/H] and
[$\alpha$/Fe] (namely [Fe/H] = [Z/H] $-$ 0.94[$\alpha$/Fe]; Thomas,
Maraston \& Bender 2003), a trend of exactly [Fe/H] peak values of
MRGC populations with mass of their hosts is of the order of
$\Delta$[Fe/H]$\sim$ 0.1 dex (using $\Delta$[Z/H]$_2 \approx$ 0.25
dex) or smaller.

One caveat about the estimates made above from the model relations
is that any such a relation (for a given age) between the color
$(g-z)$ and metallicity, [Z/H], is assumed to be the same for
different metal mixture (i.e., of [Fe/H] and [$\alpha$/Fe] ratios)
for any [Z/H]. However, the situation may be more complicated as
discussed recently by Salaris \& Cassisi (2007) for other broad-band
colors. It is not clear (1) what an order of magnitude of possible
color difference may be just in the color $(g-z)$ for various metal
mixture at the same metallicity [Z/H], and (2) how such a color
difference depends on metallicity at a given age. Hence it is
difficult to adequately take into account such effects (if any) in
the present study.

Therefore, there are two principal alternatives. First one is the
case of no age trend among MRGC populations belonging to spheroids
of different mass. It assumes only one factor, [Fe/H] trend,
responsible for a peak color trend of the populations with galaxy
mass. Second one, corresponding to the case of age trend, implies
three contributors to the color trend: age, [$\alpha$/Fe] and [Fe/H]
systematic variations. As we show above, the color trend is likely
caused by a number of factors rather than by the only variation of
iron abundance. Up to $\sim$50\% of this trend may be due to an age
trend of the MRGC populations. In this event $\sim$30\% of the peak
color trend originate from systematic variation of alpha-element
abundance ratio, and the rest ($\sim$20\%) from variation of exactly
[Fe/H] ratios.

\section{Summary and concluding remarks}
\label{sumrem}

By summarizing and analyzing the results available on MRGC
populations and on their parent galaxies, we conclude the following.
On the one hand, the strong evidence is provided for the dependence
of peak {\it color} of the populations on mass (velocity dispersion)
of their hosts. On the other hand, it is now definitely unknown
which of the factors, metallicity or age trend, contributes more to
this dependence. The latter is assumed to be small or negligible as
compared to the former one. Strictly speaking, there is only one
more or less reliable constraint on age of MRGC populations
belonging to spheroids with mass M$_s >$ 10$^{10}$M$_\odot$.
Specifically, the populations are found to be typically older than
8-10 Gyr. That is, they are old with uncertainty of $\sim$ 4-6 Gyr.
However, as we demonstrate and argue here, there is no observational
reason to neglect systematic age variations among MRGC populations
within this age range. On the contrary, from indirect evidence the
existence of a trend of age of the populations with galaxy mass is
rather more probable than its absence. The contribution of such an
age trend to the populations' color trend can be comparable to the
contribution of metallicity trend. Therefore, the assumption that
age effects are small or negligible, and the enforced conversion of
a color-mass trend to a metallicity-mass trend may ultimately lead
either to inadequate conclusions on the processes responsible for
the formation of MRGCs or to overestimating role of some of these
processes and missing role of other ones. For this reason we
estimate possible systematic age variation between MRGC populations
belonging to spheroids in mass range of nearly two order of
magnitude. Furthermore, by relying on currently available
observational data and models, we evaluate an order of magnitude of
the respective effect of the MRGC populations' age trend on the
deduced trend of their peak metallicity with mass of their hosts. We
also show that in the case of such an age trend, the real systematic
variation of total metallicity, [Z/H], of the populations is
presumably mainly due to an age$-$[$\alpha$/Fe] correlation among
GCs. Hence systematic variation of exactly [Fe/H] ratios, another
constituent of the total metallicity, is smaller or even none.

We have recently argued (Kravtsov 2006) that the formation of both
the old MRGCs in spheroids and young/intermediate-age massive star
clusters in irregulars, as well as increased SF activity
accompanying the formation of the clusters, occur at (approximately)
the same stage of the host galaxies' chemical evolution. These
processes achieve their maximum around [Fe/H] $\sim -0.5$ dex (Z
$\sim$ Z$\odot$/3). Here we show that real range of systematic
metallicity spread, particularly expressed in [Fe/H], of MRGC
populations is likely more limited than the metallicity range
usually deduced from the populations' peak color variation. The
latter range is probably the upper limit and is related to the total
metallicity, [Z/H], rather than to [Fe/H]. It can fully be
attributed to [Fe/H] only provided that there is insignificant or no
age trend of the populations, as already mentioned above.

It would be useful to compare the metallicity range deduced from the
discussed color trend of MRGCs with those obtained using other
methods that are more capable to disentangle age and metallicity
effects. In particular, most reliable spectroscopic metallicity
estimates for the MRGCs belonging to (the same) galaxies at two
extremes of their mass range could presumably allow one to impose
more strict constraints on the metallicity trend (and therefore on
the age trend) of the MRGC populations. Other alternative might be
the use of different colors which are more sensitive to metallicity
and less sensitive to age, such as $V-K, B-K$, etc. 

\acknowledgements We thank the anonymous referee for useful comments
that have improved the manuscript.

\end{document}